\documentclass[%
 reprint,
superscriptaddress,
 amsmath,amssymb,
 aps,
prb,citeautoscript
]{revtex4-2}

\usepackage{graphicx}
\usepackage{dcolumn}
\usepackage{bm}
\usepackage{lipsum}
\usepackage{siunitx}
\usepackage{placeins}
\usepackage{amstext}
\usepackage{mathtools} 
\usepackage{amsmath,amssymb}    
\usepackage{booktabs}
\usepackage{array}
\usepackage{xcolor}
\usepackage{tabularx}
\usepackage{ulem}
\usepackage{hyperref}
\usepackage{nameref}
\usepackage{float}

\usepackage{booktabs}

\DeclareSIUnit{\rad}{rad}
\DeclareSIUnit{\deg}{deg}

\hypersetup{
        colorlinks   = true,
        citecolor    = blue,
        linkcolor    = blue}

\DeclareSIUnit\bar{bar}
\DeclareSIUnit\torr{Torr}

\makeatletter
\def\@email#1#2{%
 \endgroup
 \patchcmd{\titleblock@produce}
  {\frontmatter@RRAPformat}
  {\frontmatter@RRAPformat{\produce@RRAP{*#1\href{mailto:#2}{#2}}}\frontmatter@RRAPformat}
  {}{}
}%

\let\svthefootnote\thefootnote
\newcommand\freefootnote[1]{%
  \let\thefootnote\relax%
  \footnotetext{#1}%
  \let\thefootnote\svthefootnote%
}

\begin{document}
\title{Preserving the Josephson Coupling of Twisted Cuprate Junctions via Tailored Silicon Nitride Circuits Boards}

\author{Tommaso Confalone}
\affiliation{Leibniz Institute for Solid State and Materials Research Dresden (IFW Dresden), 01069 Dresden, Germany}
\affiliation{Institute of Applied Physics, Technische Universität Dresden, 01062 Dresden, Germany}

\author{Flavia Lo Sardo}
\affiliation{Leibniz Institute for Solid State and Materials Research Dresden (IFW Dresden), 01069 Dresden, Germany}
\affiliation{Institute of Materials Science, Technische Universität Dresden, 01062 Dresden, Germany}

\author{Domenico Montemurro}
\affiliation{Department of Physics, University of Naples Federico II, 80125 Naples, Italy}

\author{Davide Massarotti}
\affiliation{Department of Electrical Engineering and Information Technology, University of Naples Federico II, I-80125 Naples, Italy}

\author{Valerii\,M.\,Vinokur}
\affiliation{Terra Quantum AG, 9000 St.\,Gallen, Switzerland}

\author{Genda Gu}
\affiliation{Condensed Matter Physics and Materials Science Department, Brookhaven National Laboratory, Upton NY 11973, USA}

\author{Francesco Tafuri}
\affiliation{Department of Physics, University of Naples Federico II, 80125 Naples, Italy} 

\author{Kornelius Nielsch}
\affiliation{Leibniz Institute for Solid State and Materials Research Dresden (IFW Dresden), 01069 Dresden, Germany}
\affiliation{Institute of Applied Physics, Technische Universität Dresden, 01062 Dresden, Germany}
\affiliation{Institute of Materials Science, Technische Universität Dresden, 01062 Dresden, Germany}

\author{Golam Haider}
\affiliation{Leibniz Institute for Solid State and Materials Research Dresden (IFW Dresden), 01069 Dresden, Germany}

\author{Nicola Poccia}
\thanks{\url{nicola.poccia@unina.it}}
\affiliation{Leibniz Institute for Solid State and Materials Research Dresden (IFW Dresden), 01069 Dresden, Germany}
\affiliation{Department of Physics, University of Naples Federico II, 80125 Naples, Italy}


\begin{abstract}
Controlled fabrication of twisted van der Waals heterostructures is essential to unlock the full potential of moiré materials. However, achieving reproducibility remains a major challenge, particularly for air-sensitive materials such as Bi$_2$Sr$_2$CaCu$_2$O$_{8+\delta}$ (BSCCO), where it is crucial to preserve the intrinsic and delicate superconducting properties of the interface throughout the entire fabrication process. Here, we present a dry, inert and cryogenic assembly method that combines silicon nitride nanomembranes (NMBs) with pre-patterned electrodes and the cryogenic stacking technique (CST) to fabricate high-quality twisted BSCCO Josephson junctions (JJs). This protocol prevents thermal and chemical degradation during both interface formation and electrical contact integration. We also find that asymmetric membrane designs, such as a double cantilever, effectively suppress vibration-induced disorder due to wire bonding, resulting in sharp and hysteretic current–voltage characteristics. The junctions exhibit a twist-angle-dependent Josephson coupling with magnitudes comparable to the highest-performing devices reported to date, but achieved through a straightforward and versatile contact method, offering a scalable and adaptable platform for future applications. These findings highlight the importance of both interface and contact engineering in addressing reproducibility in superconducting van der Waals heterostructures.\\

\end{abstract}

\maketitle

\section{Introduction}
Since the emergence of the 2D materials field, research has advanced rapidly, driven by the ability to vertically stack atomically thin layers into van der Waals heterostructures. This approach enables the creation of materials with customized electronic, optical, and mechanical properties. A particularly exciting advancement in recent years has been the advent of 2D moiré materials, where slight lattice mismatches or twist angles between stacked layers give rise to moiré superlattices with highly tunable physical behavior \cite{Mak2022, He2021, Geim2024}. However, this outstanding potential is hindered by limited reproducibility, which remains a critical barrier to the reliable scaling and commercialization of 2D-based devices \cite{Lau2022, Boggild2024, Shuck2024}. In particular, the deterministic assembly of van der Waals heterostructures faces persistent issues such as transfer-induced inhomogeneity, interfacial contamination, and variability introduced by manual fabrication processes \cite{Castellanos-Gomez2022,Ren2025}. Recent efforts to address these limitations include the automation of the transfer process \cite{Masubuchi2018} and the fabrication of heterostructures in ultra-high vacuum environments \cite{WangW2023}.\\
The issue of reproducibility is particularly pronounced when working with air-sensitive 2D materials such as Bi$_2$Sr$_2$CaCu$_2$O$_{8+\delta}$ (BSCCO), where preserving intrinsic physical properties during processing is especially challenging. Among high temperature superconductors (HTSCs), BSCCO exhibits several practical features that make it highly relevant for fundamental studies and technological applications. It possesses a layered structure, with van der Waals forces between BiO planes, enabling mechanical exfoliation with a scotch tape down to the monolayer limit on conventional SiO$_2$/Si substrates \cite{Novoselov2005}. Remarkably, even as a monolayer, BSCCO has been reported to retain superconducting properties comparable to those of the bulk material \cite{Yu2019, Zhao2019}. Additionally, its crystal structure includes a natural stack of intrinsic Josephson junctions (IJJs) along the c-axis, where superconducting CuO$_2$ planes are separated by insulating [SrO–BiO] bilayers \cite{Kleiner1992}. Finally, the Josephson coupling is predicted to exhibit strong dependence on interlayer twist angle due to the anisotropic superconducting order parameter (SOP) \cite{Klemm1998, Bille2001, Tanaka1997}.\\
Despite these advantages, the realization of clean and coherent vertical heterostructures is hindered by several factors. The short c-axis coherence length ($<$0.1nm) \cite{Naughton1988} require atomically sharp interfaces and the high mobility of the oxygen dopants above 200\,K \cite{Fratini2010, Poccia2011, Campi2015} and the highly chemical reactivity \cite{Sandilands2010, Sandilands2014, Huang2022} require a cryogenic and inert environment during sample fabrication.  Consequently, early experiments on twisted vertical BSCCO Josephson junctions (JJs) did not reveal the anticipated strong dependence on twist angle \cite{Li1999, Latyshev2004}. It was only with the recent introduction of the cryogenic stacking technique (CST) under ultra-pure Ar atmosphere \cite{Zhao2023}, that multiple research groups succeeded in preserving interfacial superconductivity in twisted BSCCO crystals and observing an angular dependence of the Josephson coupling \cite{Martini2023, Lee2023, Patil2024, Ghosh2024, Zhu2023, WangH2023}. However, variations in both the magnitude and the specific dependence of the angular response across different studies underscore that the issue of reproducibility persists.\\
While CST has helped standardize the fabrication of fragile twisted BSCCO structures, establishing consistent and reliable electrical contacts remains a challenge \cite{Confalone2025, Batool2023}. Common strategies, such as stencil masks with thermal \cite{Zhao2023, Martini2023, Lee2023} or e-beam evaporation \cite{Patil2024, Ghosh2024}, and pre-patterned electrodes \cite{Zhu2023, WangH2023}, often involve direct metal deposition or pick-up of the JJs, which could thermally and mechanically degrade the flakes. In addition, polymers and solvents pose the risk of chemical degradation, further narrowing viable options. To address these issues, we previously developed a fabrication method based on silicon nitride nanomembranes (NMBs), which decouples electrode patterning from the handling of the active material \cite{Saggau2023, Martini2023bis}. This technique enables the realization of atomically thin BSCCO devices with superconducting properties comparable to bulk, while entirely avoiding polymers, solvents, and elevated temperatures \cite{Shokri2024}.\\
Here, we demonstrate the implementation of the CST with silicon nitride NMBs for fabricating twisted BSCCO JJs. Using the dry transfer method at cryogenic temperatures, we assembled several devices across a range of twist angles between 0$^\circ$ and 45$^\circ$, while electrical contacts were formed by directly transferring the nanomembranes onto the junctions. We found that precise membrane shaping is essential for achieving sharp current–voltage (\textit{I–V}) characteristics, as symmetric geometries lead to detrimental interfacial disorder that degrades junction performance. Using this approach, we reliably fabricated high-quality twisted BSCCO JJs that exhibit an angular dependence of the Josephson coupling along with large coupling magnitudes, comparable to the highest-performing devices reported so far.

\section{Results and discussions}

\begin{figure*}
  \includegraphics[width=\textwidth]{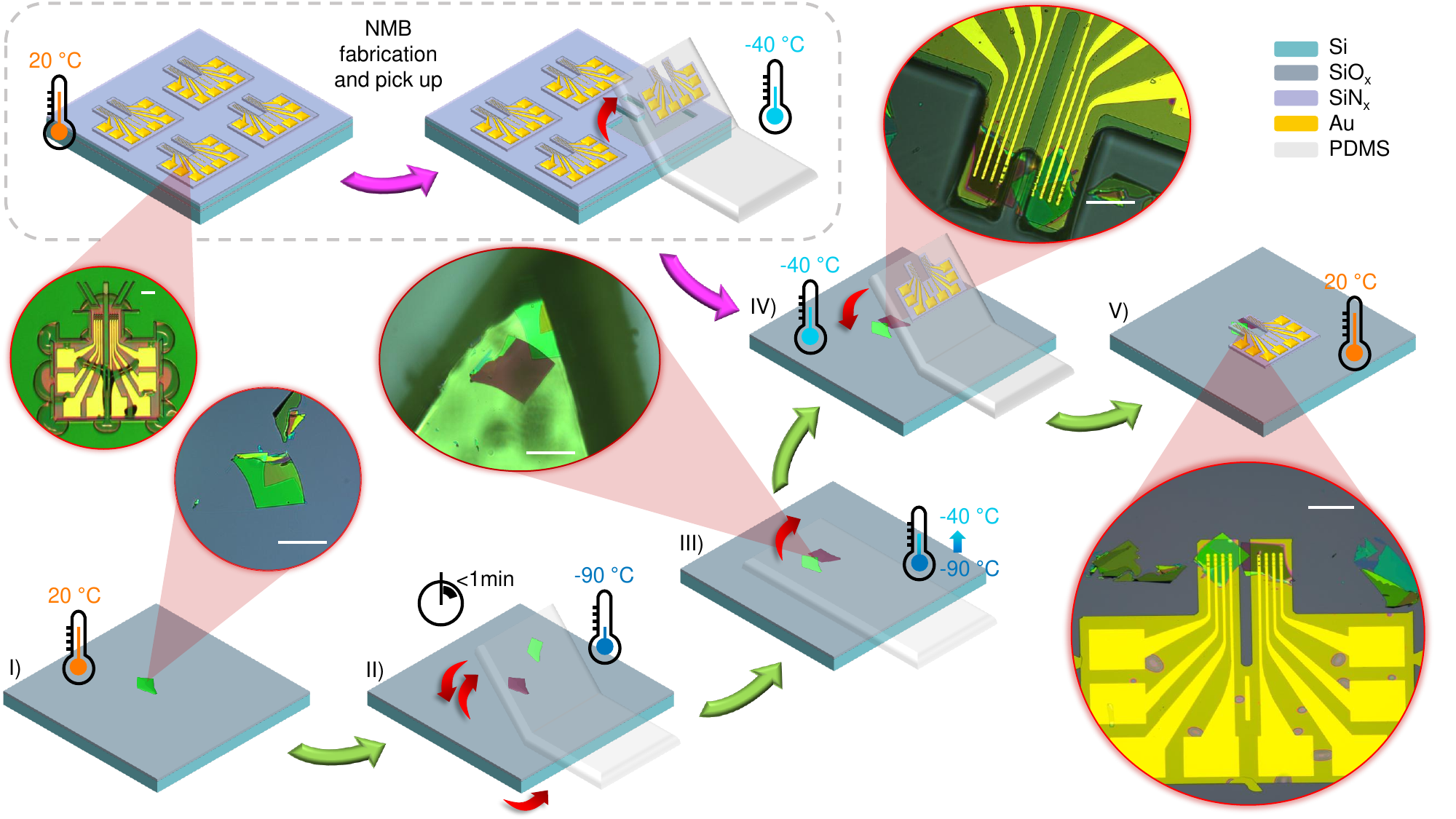}
  \caption{Schematic of the dry, cryogenic assembly process combining silicon nitride nanomembranes (NMBs) with the cryogenic stacking technique (CST) to fabricate twisted BSCCO Josephson junctions (JJs). Circular insets display optical images corresponding to each fabrication step. White scale bars represent 100\,$\mu$m. I) Mechanical exfoliation of BSCCO flakes on SiO$_2$/Si substrates. II) Cooling the stage to -90\,$^\circ$C and realization of the twisted JJ. III) Slowing warming the stage to -40\,$^\circ$C and removing the PDMS. IV) Landing a previously picked-up membrane onto the JJ. V) Warming to room temperature and removing the shaped PDMS, leaving in place the membrane on the JJ.}
  \label{fig1:fab}
\end{figure*}

\subsection{Fabrication}

The first step in fabricating devices based on air-sensitive BSCCO flakes involves the preparation of silicon nitride NMBs, which are used to dry-transfer electrodes onto the flakes under cryogenic conditions. The NMBs are fabricated in a cleanroom using standard nanofabrication techniques, following an optimized version of the process described in Ref.\,\cite{Saggau2023}. These membranes are typically prepared several days or even weeks in advance, stored in the glovebox at room temperature, and retrieved only at the moment of device assembly. Just prior to fabricating the JJs, a selected membrane is picked up using a custom-made polydimethylsiloxane (PDMS) stamp. The core idea behind the membrane design is to decouple circuit fabrication from junction assembly, allowing for electrical contact through a completely dry and low-temperature process. We emphasize that this strategy remains effective even when integrating circuit elements made from materials that would otherwise be incompatible with BSCCO's extreme chemical and thermal sensitivity \cite{Shokri2024}.\\

\subsubsection{Design and fabrication of nanomembranes}

\begin{figure*}
  \includegraphics[width=\textwidth]{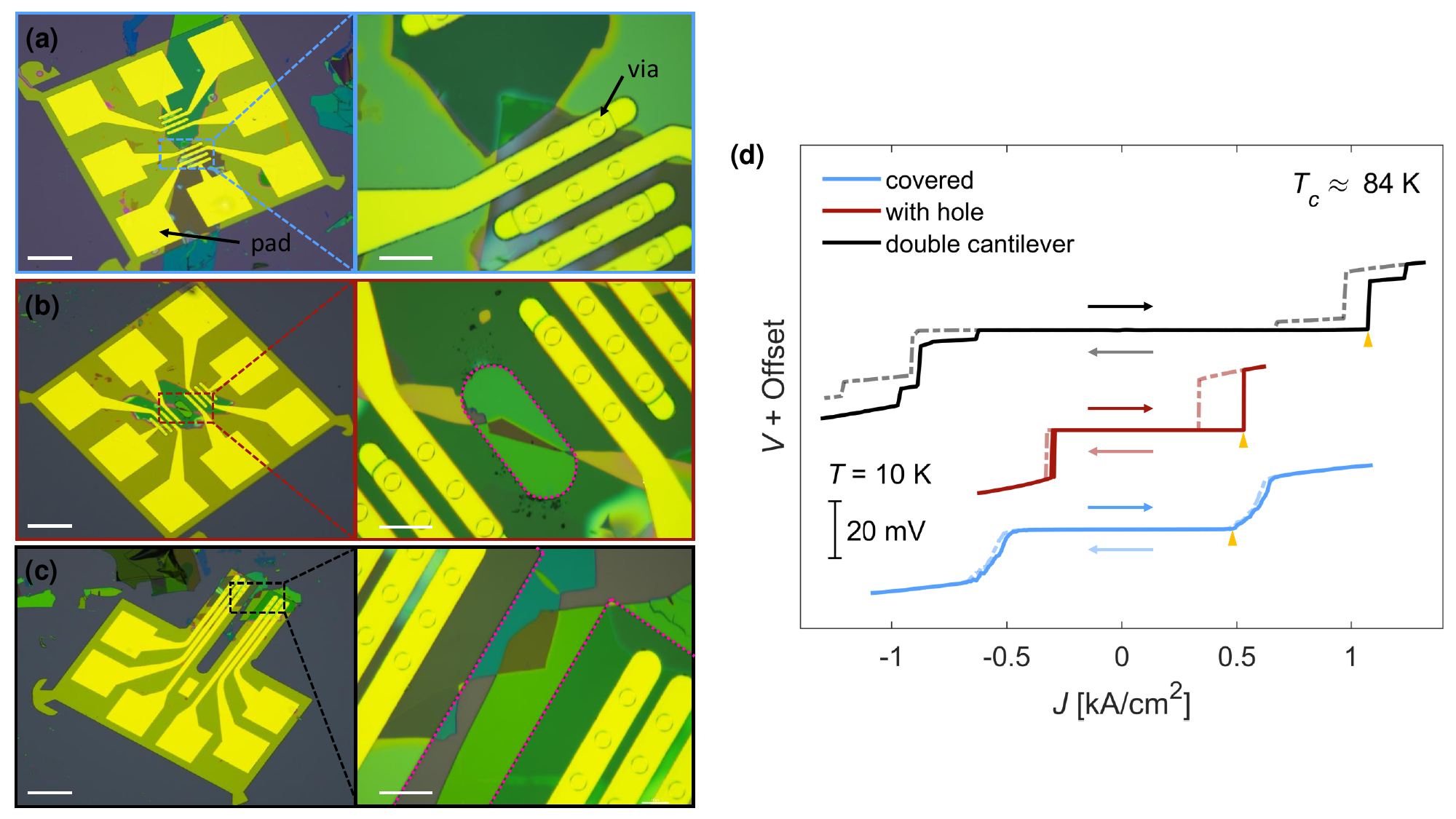}
  \caption{(a)–(c) Optical microscopy images of the three compared (covered, with hole, double cantilever) untwisted BSCCO devices, with close-up views of the junction regions. Pink dashed lines outline the edges of the nanomembranes. Scale bars represent 100\,$\mu$m for the full images and 20\,$\mu$m for the zoomed-in areas. (d) Representative current density–voltage (\textit{J–V}) characteristics measured at 10\,K for junctions fabricated with a fully covering membrane, a membrane with a hole, and the membrane with the double cantilever design. Arrows indicate the direction of the current sweep. All three samples exhibit a critical temperature of approximately $T_c \approx$ 84\,K.}
  \label{fig2:untwisted}
\end{figure*}

The fabrication of the NMBs begins by coating a Si/SiO$_2$/Si substrate with a 3.5\,nm-thick Al$_2$O$_3$ layer using atomic layer deposition (ALD). This layer serves as a protective barrier during the final XeF$_2$ etching step of the silicon sacrificial layer that make the NMBs freestanding and is removed at the end of the process.\\
The NMBs design process starts with the definition of the bottom contacts. These are patterned using optical lithography and formed by sputtering 80\,nm Au onto the deposited Al$_2$O$_3$ layer. A $\sim$\,400\,nm-thick Si$_3$N$_4$ layer is then deposited over the entire structure via chemical vapor deposition (CVD). This nitride layer forms the mechanical body of the NMBs and connects the bottom and top contact layers. Next, reactive ion etching (RIE) is used to define both trenches and vias in the Si$_3$N$_4$ layer. The trenches shape the outer perimeter of the NMB and simultaneously open a path through to the underlying silicon, while the vias expose the Au bottom contacts to allow electrical connection to the upper pads. The top contacts are then defined by optical lithography and formed by sputtering a stack of 5\,nm Cr and 80\,nm Au, establishing the connection between the lower and upper parts of the NMBs. \\
Once this electrical routing is complete, deep reactive ion etching (DRIE) is used to remove part of the silicon within the trenches, exposing the buried SiO$_2$ layer. A second 5\,nm-thick Al$_2$O$_3$ layer is deposited over the entire chip for protecting the top part of the NMBs. The access windows are then opened in this top Al$_2$O$_3$ layer via RIE to allow XeF$_2$ gas to enter and selectively etch the sacrificial silicon layer beneath, releasing the NMBs and allowing them to stand free. In the final step, both the top and bottom Al$_2$O$_3$ protective layers are removed by wet etch using Tetramethylammonium hydroxide (TMAH) solution. To prevent NMBs, collapse due to capillary forces during drying, a critical point dryer (CPD) is employed. After the CPD the NMBs can be transferred to the glovebox and are ready to be used. Details of each fabrication step can be found in the Methods section, and cross-sectional schematics of the steps are provided in Figure S1.\\

\subsubsection{Engineering PDMS stamp}
Typically, PDMS stamps are fabricated by mixing the elastomer with the curing agent and pouring the mixture on a petri dish for curing. However, to prevent any contact during the membrane landing process, we tailored the shape of the PDMS to match the geometry of the NMBs. This was achieved by using a mold, into which the PDMS is poured prior to curing, instead of the standard flat petri dish. The mold is fabricated by patterning a 50–100 \,$\mu$m-thick SU-8 photoresist layer on a standard SiO$_2$/Si substrate. We found that if the mold is thinner than 50 \,$\mu$m, the entire PDMS stamp, not just the patterned region, comes into contact with the substrate during landing, preventing control over the contact area. Details on the mold and PDMS fabrication can be found in the Methods section and optical images of the used PDMS are provided in the Supplementary Information (Figure S2). We note that by leveraging the ability to tailor both the NMB geometry and the corresponding PDMS stamp, we are able to land the membrane onto the device without touching the interface region at any stage during the fabrication of the contacts.\\

\subsubsection{Realization of twisted cuprate Josephson junctions}
We fabricated JJs using optimally doped BSCCO flakes through the CST performed entirely in an ultra-pure argon-filled glovebox. A total of seven devices were realized: three at a twist angle $\theta = 0^\circ$ and four with twist angles ranging from $0^\circ$ to $45^\circ$. The key steps of the device fabrication are sketched in Figure\,\ref{fig1:fab}. The process begins by mechanically exfoliating BSCCO single crystals onto standard SiO$_{2}$/Si substrates pretreated with oxygen plasma. Before using them, the substrates are backed at 150 $^\circ$C overnight in the glovebox to remove adsorbed water. Suitable flakes are identified under an optical microscope, and the substrate with the selected flake is then transfered onto a nitrogen-cooled stage and cooled down to approximately –90 $^\circ$C. At this temperature the crystalline integrity and the superconducting properties of the interface are preserved during junction realization. When the temperature approached –90 $^\circ$C, approximately 5 $^\circ$C above the target, the polydimethylsiloxane (PDMS) stamp is bring in contact with the surface of the flake. At this temperature, the PDMS becomes adhesive due to its proximity to the glass transition temperature ($\sim$\,–120 $^\circ$C), allowing us to cleave the flake along the BiO planes during lifting, with one portion remaining on the substrate and the other adhering to the PDMS. Subsequently, the substrate is rapidly rotated to the desired twist angle using a custom-built copper rotational stage, and the cleaved flake remained on the PDMS is aligned and restacked onto the flake on the substrate within one minute to form the JJ. The stage is then slowly warmed to –40 $^\circ$C to reduce PDMS adhesion, enabling clean removal without damaging the device. The chosen starting BSCCO flake typically exhibit lateral dimensions exceeding 100\,$\mu m$ and thicknesses in the range of 100–200\,nm. We found that with these conditions the exfoliation usually give us two cleaved flakes sufficiently thick (above 30\,nm).\\

\subsubsection{Establishing electrical contacts on heterostructures using nanomembrane} 
To establish electrical contacts a selected NMB with pre-patterned electrodes is picked up at -40\,$^\circ$C using the custom-shaped PDMS stamp that match the geometry of the NMB. After the pick up, without changing the temperature, the membrane is aligned and transferred onto the junction to realize the connection between the bottom contacts of the membrane and the device.  We found that landing the membrane at lower temperatures improve adhesion and reduces the chances of detachment from the substrate during the subsequent wire bonding steps. Finally, the entire stack is warmed to room temperature, enabling the complete removal of the PDMS and leaving the membrane in place. This low-temperature, dry transfer protocol enables the direct transfer of electrodes onto the JJ, resulting in high-quality electrical contacts characterized by an areal resistance below 35\,k$\Omega\mu m^{2}$ \cite{Saggau2023}.\\

\begin{figure*}
  \includegraphics[width=\textwidth]{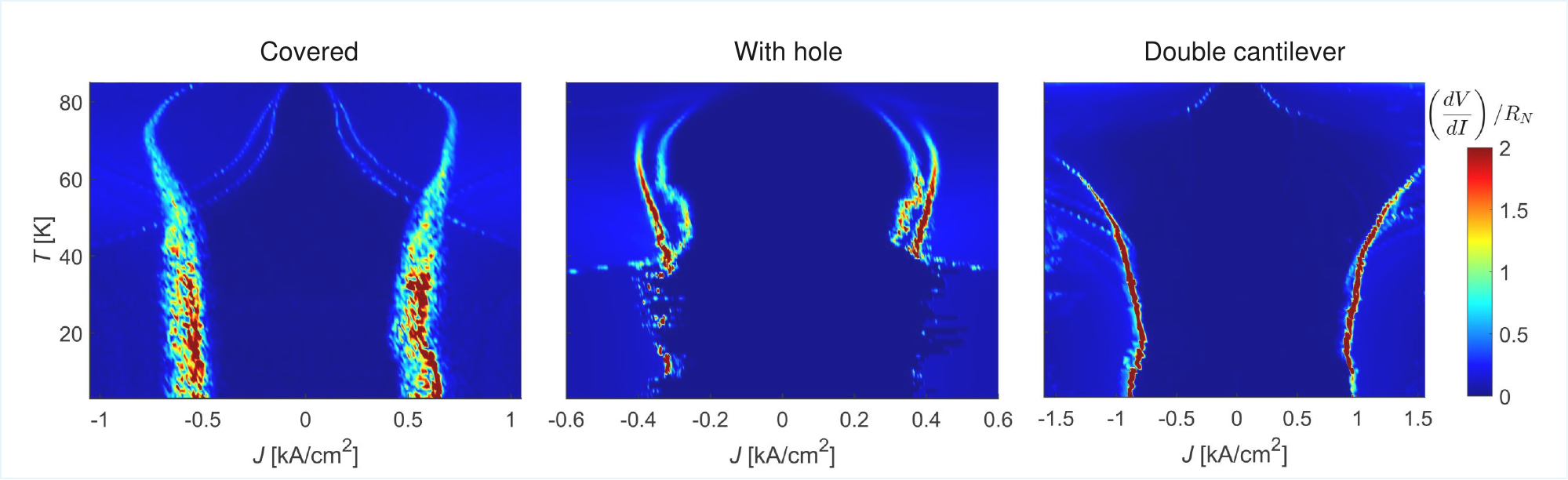}
  \caption{Normalized differential resistance ($dV/dI$)/$R_N$ as a function of bias current density $J$ and temperature $T$ from 5\,K to 85\,K of the compared untwisted JJs shown in Figures\,\ref{fig2:untwisted}(a)-(c).}
  \label{fig3:dVdI untwisted}
\end{figure*}

\subsection{Influence of nanomembrane geometry on device performance}

Using the protocol described above, we began fabricating 0$^\circ$ JJs, i.e., untwisted junctions, using three different types of NMBs, which differ solely in their geometry. The fabrication and stacking procedures are identical for all cases, with the only variation being the use of a PDMS stamp shaped to match each specific NMB during the transfer step. Optical images of the specific PDMS stamps used are provided in the Supplementary Information (Figure S2).\\
Figures\,\ref{fig2:untwisted}(a)–(c) present optical images of the three devices, with magnified views of the JJ regions. The first device (Figure\,\ref{fig2:untwisted}(a)) uses a square NMB to establish electrical contacts that completely covers the junction region. The second sample (Figure\,\ref{fig2:untwisted}(b)) employs a similar square NMB, but includes a hole (outlined with a pink dotted line) above the interface. This hole prevents the membrane from touching the interface directly during landing. In contrast, the third JJ (Figure\,\ref{fig2:untwisted}(c)) has electrical contacts realized using an NMB featuring a double-cantilever geometry. This design aims to position the contact pads (yellow squares visible in the optical images) away from the JJ area. The Supplementary Information contains optical images of the JJs taken before NMB landing (Figure S3).\\
The temperature dependence of the JJ resistance \textit{R(T)} was measured for all devices using a lock-in amplifier in a standard four-point configuration with a 1\,$\mu$A excitation current. In the normal state, all devices exhibit a linear \textit{R(T)} behavior, followed by a single sharp drop at the superconducting transition temperature \textit{T$_c$} (Figure S3), consistent with optimally doped Bi$_2$Sr$_2$CaCu$_2$O$_{8+\delta}$ \cite{Qiu2011}. In this study, the superconducting transition temperature \textit{T$_c$}  is defined as the first temperature at which the resistance drops to zero. All three JJs display a \textit{T$_c$} of approximately 84\,K, close to the bulk value of 91\,K \cite{Wen2008}, indicating a high degree of oxygen doping uniformity. This result agrees with previous studies on twisted BSCCO heterostructures \cite{Zhao2023, Martini2023, Lee2023}.\\
Figure\,\ref{fig2:untwisted}(d) compares the current density–voltage (\textit{J-V}) characteristics of the three untwisted JJs measured at a representative temperature \textit{T} = 10\,K. To facilitate comparison across devices of varying size, the current \textit{I}  is normalized by the junction area \textit{A}, yielding the current density \textit{J = I/A}. While all three devices exhibit comparable superconducting transition temperatures \textit{T$_c$}  and similar \textit{R(T)} behavior, their \textit{J-V} curves are visibly different. The fully covered device shows a critical current density \textit{J$_c$}   (yellow triangles) of approximately 0.5\,kA/cm$^2$, with a broad voltage jump of around 20\,mV at the transition from the superconducting to the normal state. This switching voltage magnitude \textit{V(J$_c$)} is consistent with the expected superconducting gap in ideal tunnel junctions \cite{Barone1982}. Additionally, the \textit{J-V} characteristic is non-hysteretic upon reversing the sweep direction of the current. In contrast, the second device, having a NMB with a hole on the junction interface area, exhibits a significantly sharper transition, while maintaining the same switching voltage \textit{V(J$_c$)} $\sim$ 20\,mV and a similar \textit{J$_c$} around 0.5\,kA/cm$^2$. Notably, this configuration exhibits hysteresis in the \textit{J-V} curve, a feature that is desirable for many applications. The third device, designed with a double-cantilever NMB to isolate the contact pads from the junction area, also displays a sharp, hysteretic \textit{J-V} transition with a 20\,mV voltage jump. Importantly, it demonstrates a higher critical current density of 1.1\,kA/cm$^2$, approaching the upper end of the range reported for IJJs in bulk BSCCO single crystals, where \textit{J$_c$} spans from 0.17 to 1.70\,kA/cm$^2$ at 10\,K depending on the number of junctions along the c-axis \cite{Irie2000}. These distinctions are not limited to a single temperature: as shown in Figure\,\ref{fig3:dVdI untwisted}, differential resistance measurements (\textit{dV/dI}) across the entire temperature range consistently reveal broader switching features in the fully covered sample compared to the sharper transitions observed in the double-cantilever device.\\\
\begin{figure*}
  \includegraphics[width=\textwidth]{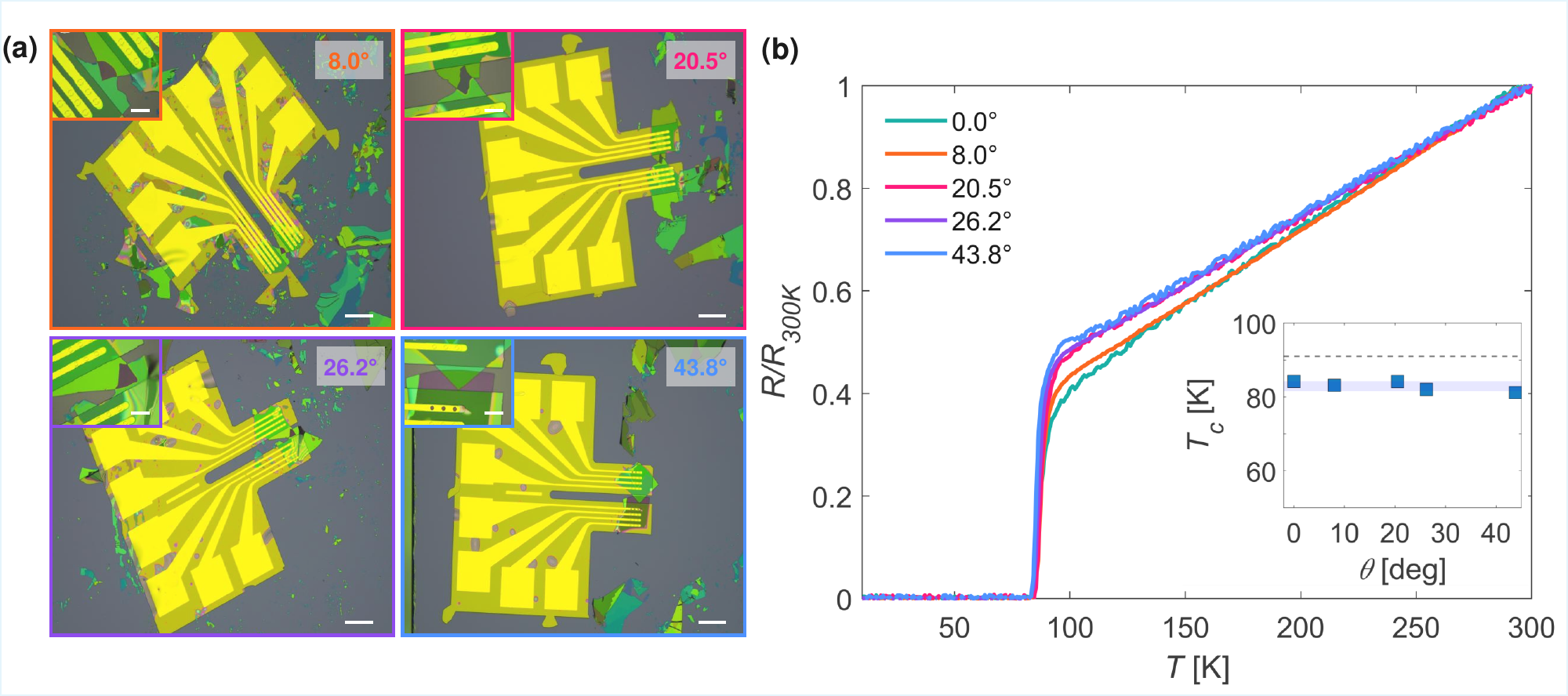}
  \caption{(a) Optical microscopy images of the four twisted devices. On the top left of each images, a zoom-in of the region of the junction while on the top right the angle of the twist. Scale bars represent 100\,$\mu$m for the full images and 20\,$\mu$m for the zoomed-in areas. (b) Temperature-dependent electrical resistance normalized at 300\,K obtained through the interface of the twisted BSCCO junctions. Inset: Angle dependence of $T_c$ of the corresponding JJs. The blue-shaded area is around the mean value of the critical temperatures and its width is two times the standard deviation. The black dashed line indicates the $T_c$ of an optimally doped bulk BSCCO crystal.}
  \label{fig4:twisted}
\end{figure*}
After carefully comparing the fabrication procedures and the realization of the NMB we attribute the improvement in the sharpness of the transition and the increase in critical current density solely to the difference in the NMB shape and NMB stacking procedure. Specifically, the double cantilever design features long arms that are connected to the main membrane pads, where bonding occurs. These extended arms naturally dampen the transmission of acoustic vibrations to the JJ interface during wire bonding, thereby reducing the risk of introducing undesirable detrimental disorder. This interpretation is supported by the observation that the JJ covered by the membrane has a symmetric design, which facilitates the propagation of vibrations, and in which the JJ area is physically connected to the NMB. While creating a hole in the NMB, as demonstrated in the second sample, already improves the JJ quality by decoupling the physical connection, it is only the asymmetric geometry of the double cantilever design that allows the JJ to achieve electronic performance comparable to IJJs in single-crystal BSCCO.\\

\subsection{twisted BSCCO Josephson junctions}
\begin{figure*}
  \includegraphics[width=\textwidth]{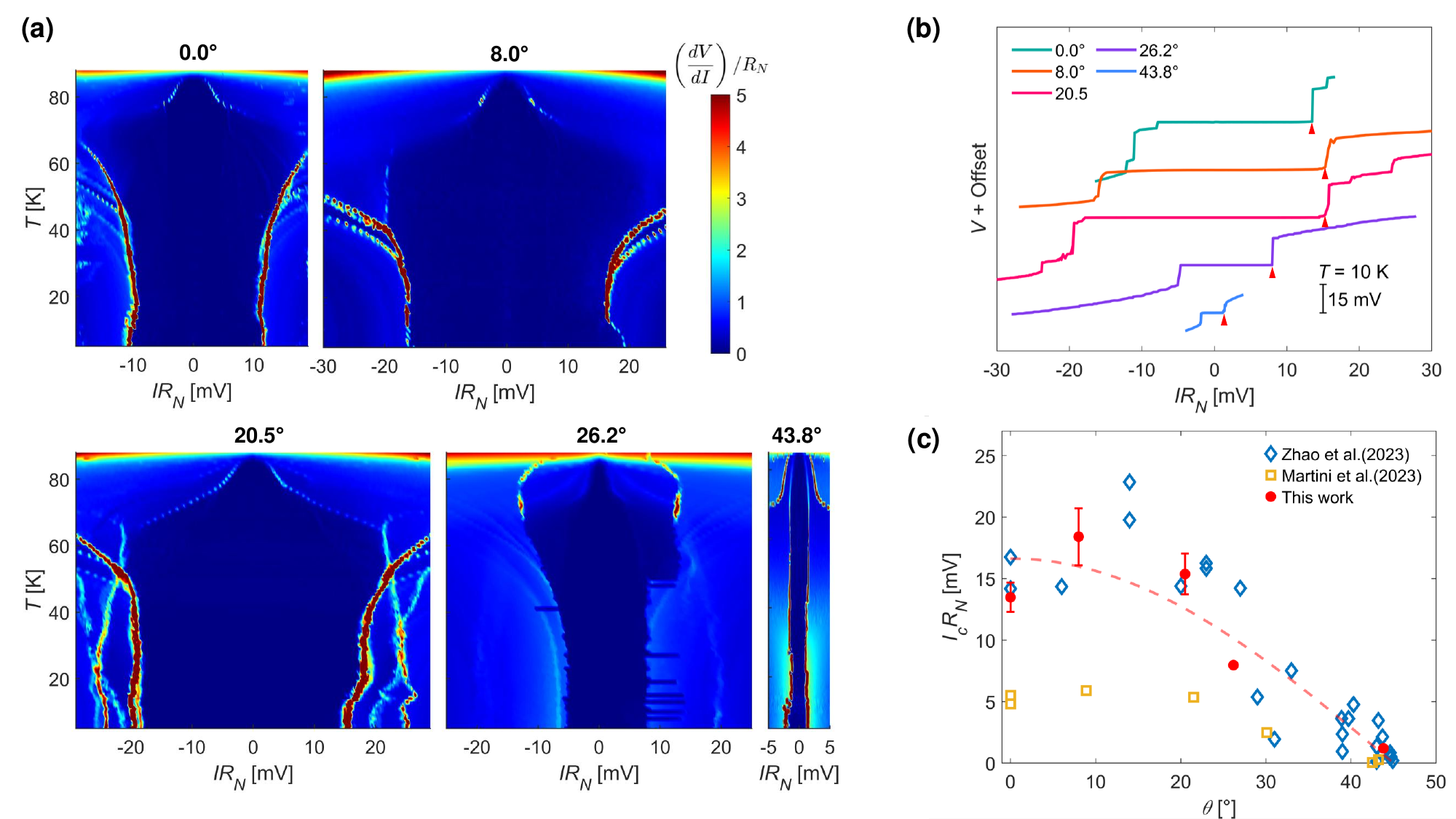}
  \caption{(a) Normalized differential resistance ($dV/dI$)/$R_N$ as a function of normalized bias current $IR_N$ and temperature $T$ from 5\,K to 88\,K of the twisted JJs shown in Figure\,\ref{fig4:twisted}(a). The twist angle is displayed on top of each color plot. The current is swept from negative to positive bias. (b) Normalized bias current–voltage ($IR_N-V$) characteristics for all JJs at 10\,K. Each curve is shifted along the y-axis for better visualization. the red triangles highlight the position of the $I_cR_N$ value were the voltage jumps occur in the $IR_N-V$ curve. (c) Comparison of the angular dependence of $I_cR_N$ between this work (10\,K), the work of Zhao et al. (12\,K) and Martini et al. (5\,K). The red dashed line follows the $cos(2\theta)$ curve, which is the expected angular dependence in first approximation for tunneling between $d-$wave superconductors. The error bars show the uncertainty on the value of $R_N$.}
  \label{fig5:angle}
\end{figure*}

After confirming the quality of the 0$^\circ$ JJ, four additional twisted JJs with angles between 0$^\circ$ and 45$^\circ$ were fabricated. Electrical contacts were made using NMBs with the double-cantilever design. The twist angle was measured using an optical microscope. Figure\,\ref{fig4:twisted}(a) display optical images of the twisted devices, with insets showing zoomed-in views of the JJ region. Except for the bottom flake of the $\theta = 43.8^\circ$ sample, all the flakes forming the JJs exhibit a similar optical color contrast, corresponding to a thickness range of 80–110\,nm. Figure\,\ref{fig4:twisted}(b) shows the resistance-temperature \textit{R(T)} trends of the twisted devices, each normalized to its resistance at 300\,K for direct comparison. As with the untwisted JJs, the devices exhibit linear behavior in the normal state and a sharp drop at the critical temperature \textit{T$_c$} . For the twisted devices, \textit{T$_c$}  varies from 81 K ($\theta = 43.8^\circ$) to 84\,K ($\theta = 20.5^\circ$), with a mean $\overline{T_c}$ = 83\,K (Inset Figure\,\ref{fig4:twisted}(b)).\\
For all devices, the current–voltage characteristics and the differential resistance across the interface are simultaneously measured as a function of temperature \textit{T}. To facilitate comparison of the transport data, the bias current \textit{I} is normalized by the normal resistance \textit{R$_N$} of the JJ. Since both the critical current \textit{I$_c$} and the inverse normal resistance \textit{R$_{N}^{-1}$}  scale linearly with the JJ area, their product is independent of the area. We define \textit{R$_{N}$} as the slope of the \textit{I–V} curve in the normal region at high bias, measured 2\,K below \textit{T$_c$}. This value is used for all temperatures, as it exhibits negligible temperature dependence, consistent with observations from previous studies \cite{Zhao2023, Martini2023}. Figure\,\ref{fig5:angle}(a) illustrates the normalized differential resistance \textit{$\frac{dV}{dI}/R_N$} for the twisted JJs as a function of temperature \textit{T} and the normalized bias current \textit{I$R_N$}, swept from negative to positive values. Intense red spots in the images represent peaks in the differential resistance, where the interface first switches from the normal to the superconducting state (for \textit{IR$_N$} $<$ 0) and then from the superconducting to the normal state (for \textit{IR$_N$} $>$ 0). The bias current at which the switch from the superconducting to the normal state occurs corresponds exactly to the value of \textit{I$_c$R$_N$}. As can be seen in Figure\,\ref{fig5:angle}(a), the normalized differential resistance is not symmetric around \textit{IR$_N$} = 0, and this asymmetry is present for all devices. Furthermore, several additional peaks, apart from the inner ones representing the switching, can be identified. These additional peaks can originate from different physical mechanism depending on the heterostructure considered. For instance, in proximitized devices, multiple peaks are typically associated to the superconducting gap and the proximity-induced superconducting gap \cite{Huang2018, Bi2024, Gu2023}. In contrast, in our devices, possible explanations for these additional peaks include phonon-assisted tunneling processes \cite{Schlenga1998}, subgap structures \cite{Schlenga1996, Itoh1997}, or IJJs intruding the current path between the voltage leads \cite{Kim1999}.\\
To clearly highlight the \textit{I$_c$R$_N$} value for each junction, Figure\,\ref{fig5:angle}(b) shows representative \textit{I–V} curves measured at \textit{T} = 10\, K. The \textit{I$_c$R$_N$} is defined as the product of the bias current at which the voltage jump occurs (indicated by red triangles) and the normal-state resistance, corresponding to the inner peaks in the differential resistance. This value typically serves as an estimate of the junction’s characteristic energy scale, with a larger \textit{I$_c$R$_N$} generally being desirable for applications that require high-performance superconducting devices \cite{Tafuri2019, Carillo2008}. Interestingly, it does not vanish approaching $\theta = 45^\circ$, as would be anticipated within a first-order approximation of tunneling between d-wave superconductors. Such a deviation aligns with previous reports \cite{Zhao2023, Martini2023, Lee2023, Patil2024, Ghosh2024} and may be attributed to co-tunneling processes of Cooper pairs \cite{Volkov2025} or to an enhanced contribution arising from inhomogeneities induced by electronic nematicity \cite{Yuan2023}.\\
Figure\,\ref{fig5:angle}(c) compares the \textit{I$_c$R$_N$} values obtained in this work as a function of twist angle with those reported by Zhao et al. (2023) and Martini et al. (2023). In all cases, the JJs were fabricated using the same cryogenic stacking method. The primary differences lie in the strategies adopted for establishing electrical contact. Zhao et al. utilized stencil mask evaporation on a –30\,$^\circ$C cold stage within a deposition chamber maintained at a base pressure of 10$^{-8}$\,mbar. Martini et al. additionally encapsulated their junctions with hexagonal boron nitride (hBN) to preserve the interface, employing the same stencil mask technique under a base pressure of 10$^{-6}$\,mbar. Here, we transfer pre-patterned electrodes on NMBs directly onto the junctions in a cold and dry manner inside an argon-filled glovebox, where the moisture level is around 10\,ppb. Despite operating at a significantly higher base pressure, our approach closely match the results reported by Zhao et al., underscoring the critical role of precise and contamination-free contact formation for reproducible and high-performance device fabrication.\\

\section{Conclusions}
In summary, we have demonstrated a robust and reliable strategy for fabricating high-quality twisted BSCCO Josephson junctions by integrating cryogenic stacking with a silicon nitride membrane-based transfer technique. This approach enables the dry, low-temperature deposition of prepatterned electrodes across twist angles ranging from 0$^\circ$ to 45$^\circ$. We find that membrane geometry plays a pivotal role in determining the junction quality. Asymmetric designs, particularly the double cantilever structure, effectively mitigate wire bonding-induced disorder, leading to sharp current–voltage characteristics with pronounced hysteresis, both of which are essential features for many superconducting device applications. Our method preserves the integrity of delicate interfaces and achieves Josephson coupling strengths comparable to the highest reported to date. These results highlight the importance of both interface and low-resistance electrical contacts engineering in enhancing the performance and reproducibility of van der Waals superconducting heterostructures, paving the way for future scalable quantum device architectures.

\section{Methods}

\textit{PDMS fabrication}: Custom polydimethylsiloxane (PDMS) stamps are fabricated using the Sylgard 184 silicone elastomer kit, following a 10:1 weight ratio of base to curing agent. The components are manually mixed for 12\,min, taking care to minimize air entrapment. The resulting mixture is then poured into the center of a Petri dish and placed in a desiccator for up to 30\,min to remove trapped air bubbles. Subsequently, the PDMS is cured by baking the Petri dish at 60\,$^\circ$C for 18\,h. After curing, the PDMS is cut into a diamond shape and transferred onto a glass slide heated at 150\,$^\circ$C for 5\,min. More details can be found in the Supplementary Information of Ref.\,\cite{Zhao2023}. To shape the PDMS according the membrane geometry the mixture is instead poured into an SU-8 mold, previously realized in cleanroom. The subsequent steps then follow the same procedure as described above. To fabricate the mold, SU-8 50 photoresist (Kayaku Advanced Materials) is spin-coated at 3000\,rpm for 45\,s onto a 1x1\,cm$^{2}$ SiO$_2$/Si substrate. The coated substrate is then subjected to a two-step soft bake: first at 65\,$^\circ$C for 10\,min, followed by 95\,$^\circ$C for 30\,min, with a 30\,min temperature ramp between each step. Next, the resist is patterned using a maskless aligner (Heidelberg MLA 100) with an exposure dose of 1500\,mJ/cm$^{2}$, followed by a post-exposure bake at 65\,$^\circ$C for 1\,min and 95\,$^\circ$C for 10\,min, using the same ramping as before. Finally, the resist is developed in mr-Dev 600 developer for 2\,min to obtain the desired mold structure.\\

\textit{Nanomembranes}: Si/SiO$_2$/Si (2\,$\mu$m/1\,$\mu$m/550\,$\mu$m) wafers (Ultrasil) are diced into 1x1cm$^2$ substrates. The substrates are cleaned by ultrasonication in acetone for 10\,min, transferred to isopropanol, dried with nitrogen, and then exposed to an oxygen plasma (50\,W) for 5\,min.\\
Al$_2$O$_3$ layers are deposited by atomic layer deposition using an Arradiance GEMstar system at 280\,$^\circ$C, with H$_2$O and trimethylaluminum as precursors.\\
For positive lithography (used in etching steps), AZ5214E (Merck Performance Materials) photoresist is spin-coated at 4500\,rpm for 40\,s, prebaked at 110\,$^\circ$C for 60\,s, exposed using a maskless aligner (Heidelberg MLA 100) at a dose of 200 \,mJ/cm$^{2}$, and developed for 60\,s in AZ 726 MIF. For negative lithography (used in liftoff steps), the same resist is applied and prebaked under identical conditions, then exposed with a dose of 10\,mJ/cm$^{2}$, post-baked at 120\,$^\circ$C for 60\,s, flood-exposed for 45\,s, and developed for 45\,s in AZ 726 MIF. \\
SiN$_x$ is deposited using plasma-enhanced chemical vapor deposition in a Sentech Si500D system at 280\,$^\circ$C, employing SiH$_4$ and N$_2$ as precursors (50\,sccm and 80\,sccm respectively) with a plasma power of 200\,W. \\
Au and Cr are sputtered using the Torr CRC600 Series system. Liftoff is performed in remover 1165 at 65\,$^\circ$C for 30\,min.\\
SiN$_x$, Al$_2$O$_3$ and Si are etched using an Oxford Instruments PlasmaPro 100 Cobra reactive ion etching system. For SiN$_x$ etching, a gas mixture of CF$_4$ (10\,sccm), CHF$_3$ (20\,sccm), and O$_2$ (4\,sccm) is used with an ICP power of 50\,W. Al$_2$O$_3$ is etched under similar conditions but with BCl$_3$ (20\,sccm) and Cl$_2$ (10\,sccm) as the process gases. For Si etching, SF$_6$ (80\,sccm) is used at an ICP power of 450\,W and a substrate temperature of -120\,$^\circ$C.\\
To etch the silicon beneath the membranes and render them freestanding, the substrates are exposed to XeF$_2$ vapor at 3\,Torr in a gas-phase etcher (SPTS Xactix). Once released, the membranes are immersed in an alkali developer (AZ 726 MIF) for 12\,min to remove the Al$_2$O$_3$ layers. They are then rinsed in deionized water for 5\,min, followed by isopropanol for another 5\,min, and finally dried using a critical point dryer (Leica EM CPD300).\\

\textit{Electrical transport measurements}: All electronic transport measurements were performed in a 9T Quantum Design Physical Property Measurement System (PPMS) that provides a temperature range from 400\,K down to 2\,K. The temperature dependence of the resistance for all the devices was measured using a four-point configuration with a lock-in amplifier (Stanford Research 830), applying an alternating current of 1\,$\mu$A at a frequency of approximately 17\,Hz. The current–voltage (\textit{I–V}) characteristics and the differential resistance (\textit{dV/dI}) across the JJs were also measured in a four-terminal configuration simultaneously. While sweeping the dc current from negative to positive values in the range of mA with a source meter unit (Keithley 2400) a small ac current with 1\,$\mu$A amplitude and a frequency of around 31\,Hz of the lock-in amplifier (Stanford Research 830) is superimposed with a custom-made transformer. The dc and the ac voltages are then simultaneously measured to get the \textit{I–V} and \textit{dV/dI} characteristics, with a digital multimeter (Keithley 2010) and a lock-in amplifier (Stanford Research 830). More details can be found in the Supplementary Information of Ref.\,\cite{Martini2023}.\\

\noindent\textbf{Acknowledgements.} N.P. acknowledges the partial funding by the European Union (ERC-CoG, 3DCuT, 101124606), by the Deutsche Forschungsgemeinschaft (DFG 512734967, DFG 492704387, DFG 539383397 DFG 460444718, and DFG 452128813) and Terra Quantum AG. The work at BNL was supported by the US Department of Energy, office of Basic Energy Sciences, contract no. DOE-SC0012704. D.M. and F.T. acknowledge support from Project PNRR MUR No. PE0000023-NQSTI.The authors are grateful to Heiko Reith and Nicolas Perez for providing access to cleanroom and cryogenic facilities respectively. T.C. gratefully acknowledges Mickey Martini and Yejin Lee for their support during the early stages of the project. The authors thank Pavel Volkov, Jedediah H. Pixley, and Marcel Franz for insightful and stimulating discussions. \\

\noindent\textbf{Author contributions.} N.P. conceived and designed the experiment. T.C. performed the experiments and analyzed the data with the contribution of F.L.S. G.G. provided the cuprate crystals. T.C., G.H. and N.P. discussed the fabrication procedure. T.C., D.M., D.M., V.M.V., F.T., G.H. and N.P. discussed the results. T.C., G.H., K.N. and N.P. wrote the manuscript. All authors discussed the manuscript.\\

\noindent\textbf{Declaration of Interest} All authors declare no conflict of interest.\\

\noindent\textbf{Data Availability Statement} The data that support the findings of this study are available from the corresponding authors upon reasonable request.

\bibliographystyle{ieeetr}
\bibliography{bibliography}

\end{document}